\documentstyle[preprint,revtex]{aps}
\tightenlines
\begin{document}
\draft
\preprint{UMD preprint \# 92-239\\
to be submitted to Phys.Rev.D\\
July '92}
\begin{title}
Hadamard States and Adiabatic Vacua
\end{title}
\author{Kay-Thomas Pirk\thanks{Internet: kp39@umail.umd.edu}\
\thanks{Bitnet: pirk@umdhep}}
\begin{instit}
Department of Physics and Astronomy\\ University of Maryland,
College Park, MD 20742
\end{instit}
\begin{abstract}
A proof is presented that in a spatially flat Robertson-Walker
spacetime the adiabatic vacuum of scalar quantum field is a Hadamard
state only if it is of infinite order and vice versa every Hadamard
state lies in the Fock space based on an adiabatic vacuum.
\end{abstract}
\section{Introduction}
It is well documented in the literature \cite{BiDa82,NaOt85} that
the so-called Hadamard condition determining the singularity
structure of the two-point function\footnote{or more precise: of the
integral kernel defining the two-point distribution} of a quantum
field is considered to be closely related to the notion of adiabatic
vacuum on a homogeneous spacetime. Both concepts are of asymptotic
nature, one with respect to distance in spacetime, the other with
wave number in phase space: the complementarity is convincing on a
heuristical level.

Several authors \cite{Birr78,BuCF78,BuPa79} did perform explicit
calculations based on both methods and derived the identical result.
But this early work focused more on the question of rendering a
concrete physical quantity finite than to establish a one-to-one
correspondence between each term of the asymptotic series attached
to either approach. Today a different emphasis seems to prevail: the
ability to renormalize the stress-energy tensor as an expectation
value is based on the proper restriction of the class of admissible
states, not on a specific computational scheme. The crucial question
remains, how to distinguish admissible states.

The Hadamard condition is certainly the most thoroughly investigated
\cite{Kay88,KaWa91} criterion for this purpose. It allows us to
separate the state-dependent information in the two-point function
of the quantum field from the purely geometrical terms that
implement the causal structure of the theory. In a linear quantum
theory this feature of the two-point function is sufficient to
establish a state as physically acceptable.

Unfortunately, complying with the Hadamard condition alone does not
suffice to qualify an arbitrary two-point function as an expectation
value of the product of two quantum field operators at different
points in spacetime. The two-point function has to meet certain
positivity requirements that result by the definition of a quantum
state as an positive functional on the algebra of observables
\cite{NaOt85,GoKa89}.

In practice the real and symmetric part of a quantal two-point
function, which contains the singularity, is obtained from a formal
integral with measure $d\mu(k)$:
\begin{equation}\label{formalint}
G^{(1)}\left(x_0,x_1\right)=\Re\int
d\mu\left(k\right)\,\psi_k^*\left(x_0 \right)\psi_k\left(x_1\right)
\end{equation}
where the functions $\psi_k$ are normalized solutions of
the linear scalar wave equation
\begin{equation}
\left[-\nabla_\mu\nabla^\mu+{\cal
V}\left(x\right)\right]\psi_k\left(x\right)=0
\end{equation}
and serve as an orthonormal basis of the ``one-particle'' Hilbert
space \cite{Wald}. This procedure ensures positivity but not
consistency with the Hadamard condition. The latter calls for a
specific but in general by no means obvious specification of the
``one-particle'' Hilbert space: the asymptotic Hadamard series has
to be transcribed into an asymptotic series for the initial
conditions on the functions $\psi_k$.

In this paper I will address this translation problem only in its
simplest form: I assume a Robertson-Walker (RW) spacetime with flat
spacelike hypersurfaces and the line-element
\begin{equation}
ds^2=a^2\left(\eta\right)\left(-d\eta^2+d\vec x^2\right),
\end{equation}
where $a$ is a $C^\infty$-function on an open interval containing
the point $\eta_i$.

In this simplified setting there exists a constructive, physically
motivated \cite{Park66,Park68} and mathematically sound
\cite{LuRo90} prescription how to specify the Cauchy data for
$\psi_k$. The states selected by the procedure are called {\em
adiabatic}. I will show that the Hadamard condition and the
adiabatic prescription are in fact equivalent concerning the class
of states they admit. Moreover, the Hadamard condition is as
specific about the initial conditions on $\psi_k$ as an asymptotic
criterion could be. In technical terms: If $\{\psi_k\}$ is a basis
suitable for the Hadamard condition, then the set ${\tilde\psi_k}$
with
\begin{equation}
\tilde \psi_k = \alpha_k\psi_k+\beta_k\psi_k^*
\end{equation}
and $\vert\alpha_k\vert^2-\vert\beta_k\vert^2=1$ is suitable as well
if and only if $\lim_{k\to\infty}k^n\beta_k=0$ for all $n$.
\section{Hadamard Series and Robertson-Walker spacetime}
The precise statement of the Hadamard condition is intricate in its
mathematical detail \cite{KaWa91}. I use the familiar yet heuristic
formulation in the tradition of DeWitt and Brehme
\cite{DeBr60,Tada87}: There exits an asymptotic expansion of the
symmetric two-point function
\begin{equation}
G^{\left(1\right)}\left(x_0,x_1\right)
=\left<\Psi\right\vert\left\{\hat\phi(x_0),\hat\phi(x_1)\right\}
\left\vert\Psi\right>\asymp{\Delta^{1/2}\over8\pi^2}\left({2\over
\sigma}+v\ln\sigma+w\right)
\end{equation}
near the coincidence limit $x_1\to x_0$.\footnote{\label{global} For
technical reasons \cite{Kay88,GoKa89} this local condition has to be
supplemented by a global one \cite{NaOt85}: The two-point function
$G^{\left(1\right)}\left(x_0,x_1\right)$ is finite whenever $x_1\neq
x_0$.} The biscalar $\sigma(x_0,x_1)$ denotes half of the square of
the geodesic interval between $x_0$ and $x_1$; $\Delta(x_0,x_1)$ is
the biscalar form of the Van-Vleck-Morette determinant. The
remaining symbols $v$ and $w$ represent biscalar functions with
smooth behaviour when $x_1$ approaches $x_0$. It is well known that
geometry and $\cal V$ alone define $v(x_0,x_1)$ by recursion
relations \cite{DeBr60,AdLN77}, whereas $w(x_0,x_1)$ contains the
information about the quantum state.

The Hadamard condition is important in part because no reference to
coordinates or peculiar symmetries of the spacetime has to be made.
However, in order to compare this condition with the notion of an
adiabatic vacuum in a RW spacetime, we have to reformulate the
Hadamard series in local coordinates:\footnote{I assume further that
the quantum state itself is homogenous and isotropic.}

\begin{mathletters}\label{hsrw}
\begin{equation}
G^{(1)}_{\scriptscriptstyle RW}\left(x_0,x_1\right)=
G^{\left(1\right)}_{\scriptscriptstyle RW}
\left(\eta_0,\eta_1,r\right)\asymp
{1\over8\pi^2a_0a_1}\left({4\epsilon\over\lambda^2}+V\ln\lambda^2
+W\right).
\end{equation}
Here we use the abbreviations
\begin{eqnarray}
\tau&:=&\eta_1-\eta_0\hphantom{\surd||},\qquad r:=\left|\vec x_1-
\vec x_0\right|,\\ \lambda&:=&\sqrt{\left|r^2-\tau^2\right|},\qquad
\epsilon:=
\mbox{\rm sign}\left(r^2-\tau^2\right),
\end{eqnarray}
\end{mathletters}
and $a_j=a(\eta_j)$. The correspondence between $v$ and $V$ or $w$
and $W$ is then given by
\begin{mathletters}
\begin{eqnarray}
v\left(\eta_0,\eta_1,r\right)&=&{\Delta^{-1/2}\over a_0a_1}V,\\
w\left(\eta_0,\eta_1,r\right)&=&\Delta^{-1/2}\left\{{4\over a_0a_1
\epsilon\lambda^2}-2{\Delta^{1/2}\over\sigma}+{V\over a_0a_1}\ln{
\lambda^2\over\left\vert\sigma\right\vert}+{W\over a_0a_1}\right\}.
\label{wexp1}
\end{eqnarray}
\end{mathletters}
The difference $\surd\Delta/(2\sigma)-1/(a_0a_1\epsilon\lambda^2)$
is a smooth function of the coordinates due to the formula
$\nabla_\mu\nabla^\mu(\surd\Delta/\sigma)=(\nabla_\mu\nabla^\mu
\surd\Delta)/\sigma$.
\section{Adiabatic Modes}
\subsection{Definition and Existence}
If we exploit the symmetry of the RW spacetime, we can express
\cite{BiDa82}the formal integral from Eq.(\ref{formalint}) as
\begin{equation}\label{rwint}
G^{\left(1\right)}_{\scriptscriptstyle RW}\left(\eta_0,\eta_1,r
\right)={1\over a_0a_1\pi^2}\int_0^\infty{\sin kr\over kr}\Re\left
\{\chi_k\left(\eta_1\right)\chi_k^*\left(\eta_0\right)\right\}k^2
dk.
\end{equation}
The modes $\chi_k(\eta)$ satisfy the
following set of equations:
\begin{mathletters}
\begin{eqnarray}
\chi_k''\left(\eta\right)+\Omega_k^2\left(\eta\right)\chi_k &=&0,\\
\Omega^2_k&=&k^2+M^2\left(\eta\right),\\ M^2\left(\eta\right)&=
&a^2\left({\cal V}-{a''\over a^3}\right)
\end{eqnarray}
and the  normalization constraint is
\begin{equation}\label{wron}
\chi_k\left(\eta\right)\chi_k'\left(\eta\right)^*-\chi_k\left(
\eta\right)^*\chi_k'\left(\eta\right)=i,
\end{equation}
\end{mathletters}
where the prime indicates differentiation with respect to $\eta$.

An adiabatic mode of $n$-th order, $\stackrel{n}{\chi}_k$, is a
specific kind of mode, defined by
\begin{mathletters}\label{zkdef}
\begin{equation}
\stackrel{n}{\chi}_k\left(\eta\right)={\exp
\left(-i\int^\eta_{\eta_i}{\stackrel{n}{W}}_k d\bar\eta\right)\over
\sqrt{2{\stackrel{n}{W}}_k}}\stackrel{n}{\zeta}_k\left(\eta\right),
\end{equation}
with the intial conditions $\stackrel{n}{\zeta}_k\left(\eta_i\right)
=1$ and $\stackrel{n}{\zeta}_k'\left(\eta_i\right)=0$, and the
iteration scheme
\begin{eqnarray}
\stackrel{0}{W}_k^2&=&\Omega_k^2,\\
\stackrel{n+1}{W}_k^2&=&\Omega_k^2
-{\cal A}\left\lbrack{\stackrel{n}{W}}_k\right\rbrack=
\Omega_k^2-{1\over2}
\left({{\stackrel{n}{W}}_k''\over{\stackrel{n}{W}}_k}-{3\over2}
\left({{\stackrel{n}{W}}_k'\over{\stackrel{n}{W}}_k}\right)^2
\right).
\end{eqnarray}
\end{mathletters}

The definition is feasible only if $\stackrel{n}{W}_k^2>0$ holds
throughout the dynamical evolution. But if
$\vert\partial^m_{\eta\ldots\eta}M^2\vert$ is bounded for all
natural numbers $m$, then $\stackrel{n}{W}_k^2>0$ is true beyond a
certain threshold ${k_{\scriptscriptstyle\asymp}}(n)$ for the wave
number $k$.

Analyzing the asymptotic behaviour of ${\stackrel{n}{W}}_k$ as done
in a recent paper by L\"uders and Roberts \cite{LuRo90} one can
infer \cite{Pirk91} that for all
$k\geq{k_{\scriptscriptstyle\asymp}}(n)$:
\begin{enumerate}
\item $\stackrel{n}{W}_k/k$
 is represented by a convergent power series in $k^{-2}$,
\item $O\left(1-\stackrel{n}{\zeta}_k\right)
=O\left(k^{-2\left(n+1\right)}\right)$ and
\item  the expansion
\end{enumerate}
\begin{equation}\label{chiasymp}
\stackrel{n}{\chi}_k\left(\eta\right)={e^{-ik\left(\eta-\eta_i
\right)}\over\sqrt{2k}}\left(t_0\left(\eta\right)+{t_1\left(
\eta\right)\over k}+\ldots+{t_{2n}\left(\eta\right)
\over k^{2n}}+O\left(k^{-\left(2n+1\right)}\right)\right)
\end{equation}
\phantom{3.}\hskip\labelsep is valid.

As a consequence the mode product can be expanded
\begin{mathletters}
\begin{eqnarray}\label{wkexpand}
\stackrel{n}{\chi}_k\left(\eta_1\right)\stackrel{n}{\chi}_k^*
\left(\eta_0\right)&=&
{\exp\left(-i\int_{\eta_0}^{\eta_1}{\stackrel{n}{W}}_k d\bar\eta
\right)\over2\sqrt{{\stackrel{n}{W}}_k\left(\eta_1\right){
\stackrel{n}{W}}_k\left(\eta_0\right)}}\stackrel{n}{\zeta}_k
\left(\eta_0\right)\stackrel{n}{\zeta}_k\left(\eta_1\right)
\nonumber\\
&=&{e^{-ik\tau}\over2k}\left(A_0+{A_1\over k}+\ldots+
 {A_{2n}\over k^{2n}}+O\left(k^{-2\left(n+1\right)}\right)\right)
\end{eqnarray}
 with
\begin{equation}\label{chicoeff}
A_m\left(\eta_1,\eta_0\right)=\sum_{n=0}^mt_n\left(\eta_1\right)
t_{m-n}^*\left(\eta_0\right).
\end{equation}
\end{mathletters}
The coefficients $A_m$ are independent of $\eta_i$; they inherit
from the expansion of $\stackrel{n}{W}_k$ the symmetries
\begin{mathletters}\label{symmet}
\begin{eqnarray}
\Im\left\{A_{2m}\right\}=\Re\left\{A_{2m+1}\right\}&=&0,\\A_{2m}
\left(\eta_0,\eta_1\right)-A_{2m}\left(\eta_1,\eta_0\right)&=&A_%
{2m+1}\left(\eta_0,\eta_1\right)+A_{2m+1}\left(\eta_1,\eta_0
\right)=0.
\end{eqnarray}
\end{mathletters}
In the appendix we list the first four coefficients $A_m$.

The motivation for introducing adiabatic modes is to control the
ultraviolet behaviour of the modes; the immediate application of
adiabatic modes is the calculation of the ultraviolet contribution
\begin{equation}
S\left[\chi_k\right]={1\over a_0a_1\pi^2}\int_{
{k_{\scriptscriptstyle
\asymp}}}^\infty{\sin kr\over kr}\Re\left\{\chi_k\left(\eta_1
\right)\chi_k^*\left(\eta_0\right)\right\}k^2dk
\end{equation}
to the two-point function when we assume that $\chi_k$ has the same
asymptotic behaviour as $\stackrel{n}{\chi}_k$ in {\em any\/} order
$n$.

\subsection{Integration}\label{asint}
Substituting the asymptotic expansion $(\exp(-ik\tau)/(2k))
\sum_{m=0}^\infty A_mk^{-m}$ for $\chi_k\chi_k^*$, we obtain
\begin{equation}
S\left[\chi_k\right]\asymp {1\over2\pi^2a_1a_0r}\Re\int_{{k_{
\scriptscriptstyle\asymp}}}^\infty e^{-ik\tau}\sin kr\sum_{m=
0}^\infty A_mk^{-m}\,dk.
\end{equation}

We are able to perform the integration on any term of the sum if we
tacitly regard $\tau$ shifted into the complex plane:
$\tau\to\tau+i\delta$ and $\delta>0$. This operation specifies how
$G^{\left(1\right)}\left(x_0,x_1\right)$ has to be understood as an
integral kernel of a distribution \cite{KaWa91}.

The integration yields
\begin{mathletters}\label{sintegrate}
\begin{equation}
S\left\lbrack\chi_k\right\rbrack\asymp{1\over8\pi^2a_0a_1}\left\{
{4\epsilon\over\lambda^2}+U{1\over r}\ln\left\vert{r+\tau\over r-
\tau}\right\vert+\tilde V\ln\lambda^2+\tilde W\right\}
\end{equation}
with
\begin{eqnarray}
\tilde V\left(\eta_0,\eta_1\right)&=&2\sum_{n=0}^\infty\sum_{m=0}^n
\left({2n+1\choose2m}{\left(-1\right)^{n+1}\over\left(2n+1\right)!}
A_{2n+2}\right.\nonumber\\& &\phantom{2\sum_{n=0}^\infty
\sum_{m=0}^n\Bigg(}\left.{}-{2n+1\choose2m+1}
{\left(-1\right)^n\over\left(2n+2\right)!}i\tau A_{2n+3}\right)
r^{2\left(n-m\right)}\tau^{2m},\nonumber\\ & &\\
U\left(\eta_0,\eta_1\right)&
=&2\sum_{n=0}^\infty\sum_{m=0}^n\left({2n+1\choose2m+1}
{\left(-1\right)^{n+1}\over\left(2n+1\right)!}\tau A_{2n+2}\right.
\nonumber\\ & &\phantom{2\sum_{n=0}^\infty\sum_{m=0}^n\Bigg(}
\left.{}-{2n\choose2m}{\left(-1\right)^n\over\left(2n\right)!}i
A_{2n+1}\right)r^{2\left(n-m\right)}\tau^{2m}
\end{eqnarray}
\end{mathletters}
and $\tilde W$ is a smooth function of the coordinates.
\section{Comparison}\label{comp}
The singularity structure of $S[\chi_k]$ corresponds to the Hadamard
series of Eqs.(\ref{hsrw}) only if every term in the series $U$
vanishes everywhere.

The integral $S[\chi_k]$ represents a solution to the wave equation;
in order to cancel the logarithmic terms, $U/r$ and $\tilde V$ have
to be asymptotic solutions in their own right. Therefore the
equation
\begin{equation}\label{twowave}
\left\lbrack\partial^2_{\eta_1\eta_1}-\partial^2_{rr}+M^2\right
\rbrack U=0
\end{equation}
holds and $\lim_{\eta_0\to\eta_1} U=\lim_{\eta_0\to\eta_1}
\partial_{\eta_1}U=0$ is a necessary and sufficient condition that
$U$ vanishes everywhere \cite{Frie75}. It is obvious that $U$ is odd
in $\tau$; it remains to prove is that $\partial_{\eta_1}U$ vanishes
on the the hypersurface $\tau=0$ as well.

We adopt a bracket notation $[A]$ to indicate
$\lim_{\eta_0\to\eta_1}A$ and derive from the above formula for $U$
\begin{equation} \label{reqcon}
\left[\partial_{\eta_1}U\right]=2\sum_{n=0}^\infty\left(-1
\right)^{n+1}\left(\left[A_{2n+2}\right]+i\left[\partial_{\eta_1}
A_{2n+1}\right]\right){r^{2n}\over\left(2n\right)!}.
\end{equation}
We show that $[A_{2n+2}]=[\partial_{\eta_1}A_{2n+1}]/i$ using the
normalization constraint Eq.(\ref{wron}): it follows from
\begin{mathletters}\label{ncons}
\begin{equation}
\left[\chi_k\left(\eta_1\right)\partial_{\eta_0}\chi_k^*
\left(\eta_0\right)\right]-
\left[\partial_{\eta_1}\chi_k\left(\eta_1\right)\chi_k^*
\left(\eta_0\right)\right] =i \end{equation} that
 \begin{equation}
\Im\left[\partial_{\eta_1}\chi_k\left(\eta_1\right)\chi_k^*
\left(\eta_0\right)\right]=-{1\over2}.
\end{equation}
\end{mathletters}
The imaginary part of the asymptotic series yields using the
symmetries of Eq.(\ref{symmet})
\begin{eqnarray}\label{impart}
\Im\left[\partial_{\eta_1}\chi_k\left(\eta_1\right)\chi_k^*
\left(\eta_0\right)\right]&\asymp&\Im\left[\partial_{\eta_1}
\left({e^{-ik\tau}\over2k}\sum^\infty_{m=0}{A_m\over k^m}\right)
\right]\nonumber\\&=&\Im\left\{-{i\over2}\sum^\infty_{m=0}{\left
[A_m\right]\over k^m}+{1\over2k}\sum^\infty_{m=0}{\left[
\partial_{\eta_1}A_m\right]\over k^m}\right\}\nonumber\\&=&-{1
\over2}-{1\over2}\sum^\infty_{m=1}\left(\left[A_{2m}\right]
+i\left[\partial_{\eta_1}A_{2m-1}\right]\right)k^{-2m}.
\end{eqnarray}
Both results, Eqs.(\ref{ncons}) and (\ref{impart}), are compatible
for an arbitrary $k$ only if the required condition,
Eq.(\ref{reqcon}), is met for all $A_m$-coefficients.

\section{Conclusions}
We have established the anticipated equivalence between the Hadamard
vacuum and the adiabatic vacuum of infinite order without evaluating
explicitly any expansion coefficients. Instead the asymptotic
quality of the adiabatic iteration combined with symmetries inferred
from the normalization constraint for the modes suffices to prove
that after integration the Hadamard series dressed in coordinates
will be matched term by term. It is further obvious from
Eqs.(\ref{sintegrate}) and Sec. \ref{comp} that an adiabatic vacuum
of $n$-th order leaves the difference
$G^{(1)}[\stackrel{n}{\chi}_k]-S[\stackrel{n}{\chi}_k]$ a $C^{2n-1}$
function instead of a $C^\infty$ one.

Yet the distance between both concepts so intimately intertwined is
far from being trivial. Solutions to the wave equation that
asymptotically resemble Minkowskian plane waves may be constructed
by means of Riemann normal coordinates in a normal neighbourhood in
an arbitrary smooth spacetime, but imposing the normalization
constraint as an integral over a Cauchy hypersurface, impedes our
ability to turn these solutions easily into a set of modes with the
desired asymptotic quality. In a homogeneous spacetime the
constraint collapses to a mere wronskian determinant and there is
not necessity to handle any issue global to the hypersurface: this
facilitated our comparison between the Hadamard expansion and the
adiabatic vacuum.

The global issue is hidden in the Hadamard condition by
supplementing the raw asymptotic expansion with the global
restriction (cf. Footnote \ref{global}) that the two-point functions
is finite whenever their arguments are separate. In the
Robertson-Walker spacetime there is no need for such a global
caveat: elementary estimates assure that the integral
Eq.(\ref{rwint}) if converging in the sense of a distribution as
outlined in subsection \ref{asint}, while $0<|\tau|,r<\epsilon$ and
$\epsilon>0$, it also converges if $r$ is extended over the interval
$r>0$, and, by Cauchy evolution, if $\tau$ is spread out. So Kay's
conjecture \cite{GoKa89,Kay88} that the global addition to the
Hadamard condition is superfluous provided
$G^{\left(1\right)}\left(x,x'\right)$ is derived from an actual
state, is at least true in a spacetime as symmetric as the assumed
RW spacetime.

\acknowledgements
I would like to thank G. B{\"o}rner and D. Brill for valuable
discussions and the thorough reading of the manuscript. This work is
based partially on research performed at the Max Planck Institute
for Astrophysics in pursuance of a doctoral degree. My stay at the
University of Maryland has been made possible by a Feodor Lynen
Research fellowship of the Alexander von Humboldt Foundation and the
National Science Foundation under Grant No. PHY90-20611. \pagebreak

\unletteredappendix{ Coefficients of the
Adiabatic Expansions}
\begin{eqnarray}
A_0&=&1,\qquad A_1=-{i\over2}\int_{\eta_0}^{\eta_1}M^2d\bar\eta,
\nonumber\\A_2&=&-{1\over4}\left(M^2\left(\eta_0\right)+M^2\left(
\eta_1\right)\right)-{1\over8}\left\lbrack\int_{\eta_0}^{\eta_1}
M^2d\bar\eta\right\rbrack^2,\nonumber\\
A_3&=&{i\over8}\left\{\left(M^2\left(\eta_0\right)+M^2\left(\eta_1
\right)\right)\int_{\eta_0}^{\eta_1}M^2d\bar\eta+
\int^{\eta_1}_{\eta_0}\left(M^2\right)''+M^4d\bar\eta\right.
\nonumber\\& &{}+{1\over6}\left.\left\lbrack\int_{\eta_0}^{\eta_1}
M^2d\bar\eta\right\rbrack^3\right\},\nonumber\\
A_4&=&{1\over16}\left\{{5\over2}\left(M^4\left(\eta_1\right)+M^4
\left(\eta_0\right)+{2\over5}M^2\left(\eta_1\right)M^2\left(\eta_0
\right)\right)\right.\nonumber\\
& &{}+\left(M^2\left(\eta_1\right)\right)''+\left(M^2\left(\eta_0
\right)\right)''+\left(\left(M^2\left(\eta_1\right)\right)'-
\left(M^2\left(\eta_0\right)\right)'\right)\int_{\eta_0}^{\eta_1}
M^2d\bar\eta\nonumber\\
& &{}+{1\over2}\left(M^2\left(\eta_1\right)+M^2\left(\eta_0\right)
\right)
\left\lbrack\int_{\eta_0}^{\eta_1} M^2d\bar\eta\right\rbrack^2+
\int_{\eta_0}^{\eta_1} M^2d\bar\eta\int_{\eta_0}^{\eta_1} M^4d\bar
\eta\nonumber\\& & \left. {}+{1\over24}\left\lbrack\int_{\eta_0}^{
\eta_1}M^2d\bar\eta\right\rbrack^4\right\}.
\end{eqnarray}

\end{document}